\documentstyle{article}
\begin{document}
\begin{center}

{\bf \Large Automated Trading Systems:  Developed and  Emerging Capital Markets
}
\end{center}
\hskip3in
\begin{center}
{\Large Ondrej Hudak, \footnote{\small present e-mail: hudako@mail.pvt.sk}}
\end{center}
\hskip2in
\begin{center}
Department of Corporate Finance, Faculty of Finance,
Cesta na amfiteater 1, SK - 974 01 Banska Bystrica, Slovakia
\end{center}
\begin{center}
{\Large Jana Tothova}
\end{center}

\begin{center}
Stierova 23, Kosice, SK-040 23 Slovakia
\end{center}

\newpage
\section*{Abstract}
Automated trading systems on developed and emerging capital markets are studied in this paper.
The standard for developed market is automated trading system with 40-days simple moving average. We tested it for the index SIX Industrial for 1000 and 730 trading days. The Buy and Hold trading system was 7.80 times more profitable than this etalon trading system for active trading. Simple taking of profitable standard trading
system from a developed capital market does not lead to optimal
results on the emerging capital markets. We then studied optimized standard trading system based on the
simple moving average. The parameter of optimization was the
number of weeks. We took the range from 1 to 100 weeks.
An optimal system was that with 5 weeks. This trading system
has some of its characteristics comparable with the etalon trading system on 
the NYSE Composite Index. The emerging market is more risky than the developed market. 
The profit on the emerging market is also higher.
The range of optimized system parameter is quite robust. Observed was increase
of number of trades in the range from the 21 weeks to the 25 weeks. This
indicates creation of a new optimal middle range trading system. Results of testing for liquid shares are quantitatively similar.

\newpage
\section{Introduction}
Can active management of a portfolio of securities add value?
That was the question \cite{1} which was discussed for developed
capital markets. Those capital markets have their properties,
which are studied and understood to quite a large extension
\cite{2}. There exist algorithmic prescriptions, called
automated trading systems, for trading on these markets. It means
that there exist algorithmic prescriptions generating buy and
sell signals. These algorithmic prescriptions are unambiguously
defined rules based on fundamental and technical analysis. They are
taking into account investment characteristics of an investor.
While development in prices in future is different from that in
history, reaction of buyers and sellers on period of
similar development on the market is similar. To define
these algorithmic prescriptions for the nearest period it is
necessary to test and to optimize these algorithmic prescriptions
on recent historical data. From similarity of reaction of buyers
and sellers on similar situations we obtain for using
these algorithmic prescriptions with tested and optimized
parameters consistency. They are robust. In longer periods
it is necessary to combine these algorithmic prescriptions with
reoptimalisation of them, using for example a method of shifting window.
Main methods of using automated trading systems are as analysis of
development on the capital market and as an investment strategy,
where buyers and sellers are acting after every buy or sell
signal.

This paper has the following sections: Automated Trading Systems,
Etalon - Developed Capital Market, Emerging Capital Markets, 
Automated Trading Systems and Emerging Capital Markets, and Summary.
We studied here automated trading systems on developed and emerging capital markets. Main results are the following ones.
The standard for developed markets automated trading system  with 40-days simple moving average  was used for the index SIX Industrial from the emerging slovak capital market. It was tested for 1000 and 730 trading days. It is not an optimal trading system for the emerging market due to the fact that the Buy and Hold trading system was 7.80 times more profitable than this etalon trading system for active trading. Simple taking of profitable standard trading system from a developed capital market thus does not lead to optimal results on the emerging capital markets. We then studied optimized standard trading system based on the simple moving average. The parameter of optimization was the number of weeks in the moving average. We took the range from 1 to 100 weeks. An optimal system was that with 5 weeks. This trading system then has some of its characteristics comparable with the etalon trading system 40-days simple moving average trading on the NYSE Composite Index. The emerging market is more risky than the developed market. 
The profit on the emerging market is also higher. The range of optimized system parameter is quite robust. Observed was increase of number of trades in the range of the 21 weeks to the 25 weeks. This
indicates creation of new optimal middle range trading system. Results of testing for liquid shares are quantitatively similar to the results of testing for the SIX 30 Industrial index.

\section{Automated Trading Systems}
Advantages of automated trading systems are, \cite{2}, in
automated decision process, in discipline, in higher consistency
of results with elimination of emotions, automated elimination of
loss to lower levels and automated using of every profitable
trend, and in finding every important trend.
Disadvantages are that no system is working universally good for
every conditions on the market, as the market develops its
characteristics are changing and thus we should not relay
on historical data only. It may happen that
buy and sell signals may not be realized due to some other reasons
(low liquidity, broker inability to realize the trade,
realization of the trade at a price completely different
from that signaled by the trading system, and other random events
which may lead to the situation in which buy and sell signals
cannot be realized).
There are complementary methods to lower disadvantages of the
trading system, they are based on the following of and analyzing of
several automated trading systems and indicators simultaneously
(almost all of these trading systems should give the same trading
signal and none of them should give an opposite signal), new
algorithms are looking for and tested (artificial intelligence,
etc.). To use automated trading systems it is necessary to
satisfy conditions like liquidity of the market with securities,
large set of historical data as concerning price (at least 5
years for middle period range of trading or to have at least 30
generated trades during the period which historical data cover)
\cite{3}. Normality of distribution of returns \cite{4} on
emerging and developed markets should be verified. A user of the
automated trading systems should have defined its
characteristics like investment horizon, whether he/she is
an aggressive or conservative investor. It is necessary to know rate
of returns for investment of free (not invested on the capital market) finances to banks, and to be able to
reinvest to the same securities. Effective commission fees
should be known. Stop-loss characteristics for an investment
should be known. An investor should has enough reserve funds to
survive the worst situations on the market. The automated trading
system should be defined unambiguously and should be simple.
Defining the algorithm of the trading system one should
preserve basic principles of financial modeling, \cite{5}.
There is a principle of independence, a principle of minimization
of constrains and a principle of top-down approach.

Automated trading systems should be compared with etalons of
automated trading systems. It is necessary to compare risk and
performance of these systems. System based on passive strategy
Buy and Hold (BH) is one etalon. Another etalon on developed
capital markets is automated trading system based on 40-days
moving average indicator. There are such criteria of risk for a
given trading system like the longest and the deepest fall of the
price in an open position (this is connected with reserve funds),
number of trades (commission fees), number of win trades (small
number of win trades with large profit does not mean that such a
system would be consistent in the future).

\section{Etalon - Developed Capital Market}
As an etalon for a developed capital market we will use standard
passive reference automated system Buy and Hold. Standard
active etalon automatic trading system for developed capital market 
is 40-days simple moving average system, \cite{2}. We used 19-years (January 1977 - December 1996)
period for index NYSE Composite. For 40-days simple moving
average automated trading system we obtained profit 94.44 for this period. The
etalon Buy and Hold system gives us for the same period only
85.01 points. Number of trades  for 40-days simple moving
average automated trading system  was 57. It means 3.20 trades per
year. 35 percents were profitable. The largest loss in points was 10.60
percents from the level 145.32 of the index in the year 1986.
The largest loss in percents was 22.53 percents from the level
60.24. The index of risk for this automated trading system
defined as the ratio of total profit to the largest loss has
the value 6.13.

\section{Emerging Capital Markets}
An efficient capital market
is that market in which every
security price equals to its investment value (NPV - Net Present Value) at all  times.
Presence of the (log-) normality distribution of returns on effective markets
most of the time is that point which forms a base for
analyzing returns on diversified portfolios when the holding
period is relatively short. The only correct measures
of returns and risk are mean return per a given period and the
standard deviation. It is an empirical evidence
that the (log-) normal distribution of returns is not present all
the time with the same parameters on efficient markets (trends
are changing), and  it is not true that such
a distribution of returns is present on efficient markets
as such for given every time periods. The aim of many econophysical models is to
explain both phenomena. However simple
tests of presence of normality of distribution of returns on efficient
capital markets should be done first. It is not known whether an emerging capital
market is or it is not an efficient market in the sense in which every
security price equals its investment value at all  times. In
the case of the emerging capital market testing of the presence of the
(log-) normal distribution is even more important
than in the case of the efficient capital market. It is necessary
to find the answer to the
questions: what are the correct measures of mean returns and risk
on emerging markets?
Recently we tested normality of distribution of returns on the
slovak capital market, which is an emerging capital market, within
the period of years 1993 (the starting year of capital market
trading) - 1996. We studied presence \cite{4}
of normally distributed returns on the slovak capital market from the short time point of
view as concerning main capital market indexes, industrial indexes,
and for shares. We used a simple sign-test which
was appropriately modified. As an etalon we considered the
capital market of the U.S.A. We have found that main
indexes of the slovak capital market (SAX, SEVIS 100, RMS F, RMS
I, SIX ALL, SIX 50, SIX 30 Industrial) and indexes of industries on the same market, as well
as those shares with the highest market capitalization, have
comparable values of the index measuring  presence of
normality of distribution with those values of this
index for the etalon market. On the emerging market
we have found an industrial structure as concerning presence of
the normality of returns. With increasing time and liquidity of
shares on the emerging market the presence of the normality of
returns increases. From our results it follows that the risk may
be described by the standard deviation of returns for
main indexes of the slovak capital market and for some of
industrial indexes of the same market, as well
as for those shares which have the highest market capitalization.

\section{Automated Trading Systems and Emerging Capital Markets}
As an emerging capital market we used the slovak capital market.
We tested this market. Our aim was to find its characteristics
for analysis, and for trading (where however number of trading
days is smaller for the emerging market with respect to the
period of 19 years of the etalon above). We used for testing of automated trading systems 
the methodology from \cite{3}. As an market index we used the index
SIX 30 Industrial. Further we tested a set of liquid shares from
shares traded on Bratislava Stock Exchange (where in the period
which we studied 777 different share titles were traded) and a
set of liquid shares from the RMS Slovakia over the counter market
(where in the period studied 639 different share titles  were
traded). We used two time periods for the index and shares from
the Bratislava Stock Exchange (1000 and 730 trading days) and one
time period for shares traded on the RMS Slovakia (730 trading
days, there were not 1000 trading days on this market). 
We used data from the beginning of trading
on the Bratislava Stock Exchange in 1993. End of the time period
is on 31st August 1996. The time period 730 trading days does not
contain data from the beginning of trading until the spring of
1994. We considered Long positions traded on the closing price last value in
a given week for shares and on the last value of the index for a
given week in trades only. Thus trading was on
the Closing price for shares and on the value of the index.
Number of weeks which we used to wait to confirm the signal buy
or the signal sell was 0 weeks. Thus we traded aggressively.
We considered zero commission fees. We assumed that there existed
possibility to reinvest all of the investment again into the same
share title or index.

Main results which we obtained were the following.
For the standard automated trading system with 40-days simple
moving average for the index SIX Industrial (1000 trading days)
we have found rate of return (per annum) 3.68 percents. There
were 4 trades, one of them was profitable. The largest loss was
71.87 points. The ratio of total profit to the largest loss is
1.43 . The Buy and Hold trading system was 7.80 times more profitable.
Thus we can see that simple procedure of taking profitable standard trading
system from a developed capital market and to use it on an emerging capital market
does not lead to optimal results on the emerging capital market. Or even it does not lead to acceptable
results on this market, see number of trades and number of win trades which is very low.

We then optimized this standard trading system based on the
simple moving average. The parameter of optimization was the
number of weeks. We took the range from 1 to 100 weeks for the
SIX 30 Industrial (1000 trading days trading period). An optimal
system was that with 5 weeks. It is 25 trading days. The rate of
return (per annum) was 89.59 percents. Number of trades was 13.
There were 4 win trades, e.i. 30.08 percents of trades. This
number is comparable with the etalon, the NYSE Composite Index.
Number of trades per year is 3.71 . This number is also
comparable with the NYSE Composite etalon. The largest loss was
120.30 points. The ratio of the total profit to the largest loss
is 20.63, which is better than on the etalon market. The emerging
market is however more risky than the developed market. The Buy and Hold
trading system was 0.32 times less profitable.
As  the optimal range for active trading using trading system
based on the simple moving average we have found range from 3 to 7
weeks. This range is quite robust. We have also observed increase
of number of trades in the range of the 21 weeks to the 25 weeks
using trading system based on the simple moving average. This
increase indicates creation of a new optimal middle range trading
system. Testing the same system on the same index for 730 trading
days we obtained quantitatively similar characteristics as for
1000 trading days period. Results for liquid shares are also similar quantitatively. 

\section{Discussion}
In this paper we discussed basic automated trading systems for developed and  emerging capital markets.
Can active management of a portfolio of securities add value? That question \cite{1} was discussed for developed
capital markets. Active management of portfolio uses usually automated trading systems.
They are algorithmic prescriptions generating buy and
sell signals which may add value. These rules are
taking into account investment characteristics of an investor.
We used an empirical fact known from developed capital markets that while development 
in prices in future is different from that in
history, reaction of buyers and sellers on period of
similar development on the market is similar. Thus it has sense to define
these algorithmic prescriptions for the nearest period on the market by testing and optimizing these algorithmic prescriptions on recent historical data. From similarity of reaction of buyers
and sellers on similar situations we obtain prescriptions with parameters which are consistent and robust. 
In longer periods these algorithmic prescriptions are reoptimized, using the method of shifting window.
Automated trading systems are used mainly to analyze development on the capital market and as an investment strategy,
where buyers and sellers are acting after every buy or sell signal.
In section Automated Trading Systems we discussed advantages and disadvantages of these systems.
Automated trading systems may be used when conditions like liquidity of the market with securities and
large set of historical data as concerning price are fulfilled. Normality of distribution of returns on developed capital markets is usually present. However it should be verified. It is known that this distribution is only roughly describing distribution of returns, see for example \cite{6}. We discuss properties of developed capital markets as well as of emerging capital markets. The empirical analysis of developed capital markets lead to description of distribution of return on these markets by Truncated Lévy Flight distributions. It is not known whether this distribution is present on emerging capital markets and it is difficult to verify presence of this distribution on emerging capital markets due to their specific properties. We used properties of normal distribution of returns which is most of the time present on the developed and in some time periods on emerging capital markets, as we verified this fact. A user of the
automated trading systems has its characteristics like investment horizon and whether he/she is
an aggressive or conservative investor. In test we are using in general rate
of returns for investment of free (not invested on the capital market) finances into banks. It is assumed that there is present possibility to reinvest return into the same securities. Effective commission fees
should be known. We used in our tests zero commission fees. We did not used stop-loss characteristics for trading systems because the standard 40-days moving average method on developed capital markets does not use them. It is assumed that an investor has enough reserve funds to survive the worst situations on the market. Automated trading system tested in our paper was compared with the etalon of automated trading systems, with Buy and Hold system. Another etalon on developed capital markets is automated trading system based on 40-days moving average indicator mentioned above. 
We tested this etalon for a developed capital market and for an emerging capital market. We have found that on the emerging capital market this etalon is not a good etalon. The parameter of 40 days should be changed.
Criteria of risk for a given trading system like the longest and the deepest fall of the
price in an open position, number of trades and number of win trades may be used on emerging capital markets.
We used 19-years (January 1977 - December 1996) period for the NYSE Composite Index. For 40-days simple moving
average automated trading system we obtained profit 94.44. The etalon Buy and Hold system gives us for the same period only 85.01 points. Thus active trading on developed capital market gives a value. Number of trades  for 40-days simple moving average automated trading system  was 57 during all the period. Thus 3.20 trades per
year were found. 35 percents of the trades were profitable. The largest loss was 10.60
percents from the level 145.32 of the index in the year 1986. The largest loss in percents was 22.53 percents from the level 60.24. For this automated trading system we have found the ratio of total profit to the largest loss with the value 6.13.

On the efficient capital market every security price equals to its investment value (NPV) almost all the time.
This is not true on emerging capital markets. Tests of presence of 
the (log-) normal distribution of returns on effective markets and on emerging markets show that these markets are different. On developed capital markets most of the time this presence is a base for analyzing returns on diversified portfolios when the holding period is relatively short (CAPM and similar models). Then the only correct measures of returns and risk are mean return per a given period and the standard deviation of returns. It is an empirical evidence that the (log-) normal distribution of returns on developed capital markets is not present (with 95 percent probability for a given time period, or with a similar probability) all the time. Trends are changing on developed capital markets and thus the same parameters of the normal distribution of returns on efficient markets cannot be used for longer time periods and moreover it is not true that such as a distribution of returns ((log-) normal)is present on efficient markets as such for given time periods. Simple tests of presence of normality of distribution of returns on efficient capital markets should be thus done first also. Many of indicators in technical analysis and many models in fundamental analysis are using the fact of presence of this distribution on developed capital markets (in practical sense).
On emerging capital markets security price does not equal to its investment value (NPV) at time periods which are much more frequent than those for the developed capital market. In the case of the emerging capital market 
testing of the presence of the (log-) normal distribution of returns is necessary for the same reasons as for developed capital market. Moreover it is necessary to find the answer to such questions like what are the correct measures of mean returns and risk on emerging markets? Tests of presence of normal distribution of returns on the emerging
slovak capital market within the period of years 1993 (starting year of capital market
trading) to 1996 were performed in this paper. In this period the market was certainly an emerging market, thus we do not study latter years on this market in which it is approaching developed market properties more. Results of our tests of presence of normally distributed returns, from the short time point of view, as concerning main capital market indexes, industrial indexes, and for shares for slovak capital market using a simple sign-test of hypothesis of presence of this distribution with 95 percent probability which was appropriately modified have shown that main indexes of the slovak capital market (SAX, SEVIS 100, RMS F, RMS
I, SIX ALL SIX 50, SIX 30 Industrial),  and indexes of industries of the same market, as well
as those shares with the highest market capitalization, are reaching comparable values of the index measuring  presence of normality with those values of this index for the etalon market (NYSE in USA). With increasing time and liquidity of
shares on the emerging market the presence of the normality of returns increases. From our results it follows that the risk may be described by the standard deviation of returns for main indexes of the slovak capital market and for some of
indexes for industries of the same market, as well as for those shares which have the highest market capitalization.
We did not tested presence of the TLF distribution for the slovak capital market. Thus the question whether an active management of a portfolio of securities can add value may be extended to emerging capital markets. This was done in this paper. As an emerging capital market we used the slovak capital market, as it was mentioned.
For testing we used the methodology used for testing developed capital markets, \cite{3}. We used the index
SIX 30 Industrial as an market index. We tested also a set of liquid shares from shares traded on Bratislava Stock Exchange and a set of liquid shares from the RMS Slovakia over the counter market. Two time periods for the index and shares from the Bratislava Stock Exchange (1000 and 730 trading days) and one time period for shares traded on the RMS Slovakia market (730 trading days) were studied. Data from the beginning of trading on the Bratislava Stock Exchange in 1993 were used. The time period 730 trading days does not contain data from the beginning of trading until the spring of
1994. Thus we eliminated trading days at the beginning of trading on the market when prices on the market were developing from those resulted from the coupon privatization to those resulting from trading on the emerging capital market. We considered Long positions in trades on the last Close price value in a given week for shares and on the last value of the index for a given week in trades only. Short position trading is forbidden on slovak capital market.
Trading using software was done on the Closing price for shares and on the value of the index. We traded as aggressive investors, number of weeks which we used to wait to confirm the signal buy
or the signal sell was 0 weeks. This corresponds to the etalon for active trading systems on developed capital markets which is 40-days simple moving average trading system. Zero commission fees were considered. It was assumed that there existed possibility to reinvest all of the investment returns from a given period (from the period with open Long position) again into the same share title or index. In its simplest form we did not gave money into bank when Long position was closed. This simplification also correspond to the mentioned etalon.

The standard etalon automated trading system with 40-days simple
moving average on the index SIX Industrial (1000 trading days)
lead to the rate of return (per annum) 3.68 percents. There
were 4 trades, only one of them was profitable. The largest loss was
71.87 points. The ratio of total profit to the largest loss is
1.43 . The Buy and Hold trading system was 7.80 times more profitable.
The passive trading is more profitable than the active trading using the standard etalon 
automated trading system from developed capital markets. We see that it is not possible simply to take standard trading
system etalon from a developed capital markets and to use it for emerging capital markets. It does not lead to optimal
results on the emerging capital market. It leads even to such unacceptable
results on this market as low number of trades and low number of win trades.

We then optimized this standard trading system. The parameter of optimization was the
number of weeks in the simple moving average. We took the range from 1 week to 100 weeks for the parameter 
for the SIX 30 Industrial index with 1000 trading days trading period. An optimal
system was that with 5 weeks which corresponds to 25 trading days. The rate of
return (per annum) was 89.59 percents for this optimized trading system for this period.
This is larger than the rate of returns using Buy and Hold trading system. 
Number of trades was 13. There were 4 win trades, e.i. 30.08 percents of trades. This
number is comparable with the etalon for active trading on the developed market on the NYSE Composite Index.
Number of trades per year is 3.71 . Again this number is also comparable with the NYSE Composite etalon. 
The largest loss was 120.30 points. The ratio of the total profit to the largest loss
is 20.63, which is larger than that for trading using the etalon for developed market and trading with this etalon system on the SIX 30 Industrial. The ratio is larger than that on the developed market. The emerging market is more risky than the developed market which corresponds to higher rate of returns on the emerging market than on the developed market.  The Buy and Hold trading system was 0.32 times less profitable than the trading using the optimized trading system. The optimal range of the parameter for active trading using trading system based on the optimized simple moving average is that range for which the results are near the results for the optimized system. We have found that the optimal range of the parameter is from  3 to 7 weeks with maximum for 5 weeks. This range is quite robust around the optimal range of 5 weeks. As we so above the 40-days simple moving average trading systems is unacceptable in this case.

Observed increase of the number of trades in the range of 21 weeks to 25 weeks
using trading system based on simple moving average indicates creation of a new optimal for middle range trading
system. The same system tested on the same index for 730 trading days gave quantitatively similar characteristics as for
1000 trading days period. Thus characteristics of the emerging market were similar whether we took into account the period of beginning of trading or not. Results of testing for liquid shares are also quantitatively similar to the results of testing optimized trading system on the index SIX 30 Industrial. This index contains in its portfolio liquid shares, so the observed similarity is not surprising. Automated trading systems known from developed capital markets do not have the same properties on emerging capital markets. It is necessary for emerging capital markets to test these systems again.

\section*{Acknowledgements}
Part of the results presented here was preliminary presented in the authors talk given in slovak 
on the seminar Infofinex 1997 in Banska Bystrica. The author would like to express his thanks
to the organizers of seminar for invitation. The paper was supported by the grant VEGA No. 1/0495/03.

\end{document}